# One-Pot Preparation of Mechanically Robust, Transparent, Highly Conductive and Memristive Metal-Organic Ultrathin Film

*M. Moreno-Moreno[1], J. Troyano[2], P. Ares[1], O. Castillo[3], C. A. Nijhuis[4], L. Yuan[4], P. Amo Ochoa[2], S. Delgado[2], J. Gómez-Herrero[1,5], F. Zamora[2,5,]*, and C. Gómez-Navarro[1,5,]**

[1]Departamento de Física de la Materia Condensada, Universidad Autónoma de Madrid, Madrid E-28049, Spain

[2]Departamento de Química Inorgánica and Institute for Advanced Research in Chemical Sciences (IAdChem), Universidad Autónoma de Madrid, Madrid E-28049, Spain

[3]Departamento de Química Inorgánica, Universidad del País Vasco, UPV/EHU, Apartado 644, E-48080 Bilbao, Spain

[4]Department of Chemistry, National University of Singapore, 3 Science Drive 3, Singapore, 117543. Centre for Advanced 2D Materials and Graphene Research Centre, National University of Singapore, 6 Science Drive 2, Singapore 117546, Singapore,

[5]Condensed Matter Physics Center (IFIMAC). Universidad Autónoma de Madrid, Madrid E-28049, Spain.

**The future of 2D flexible electronics relies on the preparation of conducting ultrathin films of materials with mechanical robustness and flexibility in a simple but controlled manner. In this respect, metal-organic compounds present advantages over inorganic laminar crystals owing to their structural, chemical and functional diversity. While most metal-organic compounds are usually prepared in bulk, recent work has shown that some of them are processable down to low dimensional forms. Here we report the one-pot preparation, carried out at the water-air interface, of ultrathin (down to 4 nm) films of the metal-organic compound $[Cu_2I_2(TAA)]_n$ (TAA= thioacetamide). The films are shown to be homogenous over $mm^2$ areas, smooth, highly transparent, mechanically robust and good electrical conductors with memristive behavior at low frequencies. This combination of properties, as well as the industrial availability of the two building blocks required for the preparation, demonstrates their wide range potential in future flexible and transparent electronics.**

The possibility of creating *à la carte* compounds based on modular chemistry has recently triggered enormous interest among the material science community.[1, 2] In this regard, coordination polymers (CPs), also named metal-organic polymers, consisting of metal centres linked by organic ligands, offer chemical design and high degree of crystallinity. Indeed, they represent an excellent structural option since they can also provide multi-functionalities. The main current challenges in this field rely in combining processability and desired electrical and/or optical or magnetic properties in a single material to integrate them in future devices. Up to date, most of the research involving these multi-functional materials has been focused on bulk material, however processability down to lower dimensional forms is still little developed.[3,4] Some seminal works have already demonstrated one-dimensional metal-organic polymers,[5, 6] as well as CP-nanosheets obtained by either top-down methods[7-10] or bottom-up approaches.[11-15]

In this scenario, bulk experiments have shown that the combination of cooper metal centres with sulfur-containing ligands as building blocks are able to produce multi-functional materials showing electrical conductivity together with optical emission.[16] Some of these compounds have been investigated in films showing conductivities as high as 1580 S/cm for film thickness of about 200 nm.[17] Nevertheless, future developments depend on the ability to obtain even thinner but robust films to integrate them in flexible electronics devices.[18]

Here we report planar devices based on mechanically robust ultrathin films (down to 4 nm thickness) of a copper-thioacetamide metal-organic polymer exhibiting conductivities up to 50 S/cm. The synthesis of the films is carried out at the water-air interface being a simple and up-scaled method based on inexpensive and industrially available building blocks. Langmuir–Schäfer technique allows us to deposit high coverage homogeneous films on a variety of substrates. The observation of free-standing films and nanoindentation experiments demonstrate that even the thinnest films are flexible and robust mechanical entities. Local surface potential maps recorded on *in-operando* electrical devices show that electrical conductivity is mediated by grain boundary presenting memristive characteristics at low frequencies as an added value.

As reported previously by our group,[19] the reaction between CuI and thioacetamide (TAA= thioacetamide) yields to a variety of structures that can be tailored by fine-tune of the ratio between the initial building blocks and the reaction speed modulated by the concentration of the reactants and the solvent. Thus, the slow diffusion of an acetonitrile solution of CuI and thioacetamide (2:1) into diethyl ether gives rise to the formation of single-crystals. X-ray diffraction indicates that the crystals correspond to a coordination polymer of formula $[Cu_2I_2(TAA)]_n$ that presents a notable degree of structural disorder (Figure 1a and SI1) involving the copper(I) atoms, which are distributed among three positions (positional disorder) but also a position in the coordination sphere of the metal centre which is shared by an iodide atom (I3) and the TAA ligand (substitutional disorder). This degree of disorder makes it difficult to provide a straightforward description of the crystalline structure based on the coexistence of a limited number of ordered arrangements. In fact, the amount of possible ordered models contributing to the mixture is large. In order to represent some of them, Figure 1b provides just three possible structures, which are chemically and structurally compatible with a suitable crystallographic solution. In any case, all the structural models consist of a 3D crystal structure build up by means of iodide bridged by tetrahedral copper(I) metal centres with coordination spheres formed by 4 or 3 iodides and 1 sulphur atom from the thioacetamide ligand. The coordination bond distances (Cu-I: 2.51-2.68 Å and Cu-S: 2.31-2.41 Å) are in agreement with the values typically found in the CSD database.[20] This type of disorder of soft metal centres has been reported for several ionic conductors such as α-AgI, $Ag_3SI$ and $RbAg_4I_5$ compounds among others in which the silver atoms are disordered over many available positions.[21-26] Interestingly, our good quality crystals showed low electrical conductivity ($10^{-4}$ S/cm) but the more amorphous crystals reached conductivities up to 10 S/cm (see SI2). The unusual X-ray structure found for $[Cu_2I_2(TAA)]_n$ as well as these electrical features prompted us to explore the nanoprocessability of this new material.

To this end, we prepare polymer films by direct reaction carried out at the water-air interface upon addition of 50 μL of an acetonitrile solution of the two simple building blocks, CuI and TAA in 2:1 ratio (Figure 2a). Immediately after that, we observed a transparent film on the water surface. We use Langmuir–Schäfer technique to deposit the film on a variety of substrates such as $Si/SiO_2$, fused quartz or glass (Figure 2b and c). Inspection under an optical microscope of deposited films on substrates of Si with thermally grown $SiO_2$ of thickness 300 nm (Figure 2b) shows the formation of continuous and homogeneous thin films with coverage higher than 85% through $mm^2$ scale areas. UV-visible spectroscopy of films prepared on quartz substrates (Figure 2c) revealed that the transparency of the films was > 80 % for the whole visible window (Figure 2d). X-ray analysis of the as-formed $[Cu_2I_2(TAA)]_n$ films deposited on $SiO_2$ agrees with that determined for the material prepared as a single crystal, therefore confirming its structure (Figure 2e). Although the coordination bond sustained 3D architecture, the synthetic approach determines the crystal growth to be limited to the water-air interphase affording this compound in the form of ultrathin sheets. Spectroscopic and analytical data confirm that the films show a similar structure and composition to that found in the $[Cu_2I_2(TAA)]_n$ single crystals (SI3).

Atomic Force Microscopy (AFM) images, as the one shown in panel 2g from the marked area of Figure 2f, confirm that these films mostly present thicknesses ranging from 4 to 10 nm. We also found few regions with films of thickness up to 60 nm, the film thickness can be adjusted just by optical inspection of the layers created at the air-liquid interphase during the transfer process to the substrate. Histograms of topographic heights measured in areas of ≈100 $\mu m^2$ of the films, as that depicted in Figure 2h, showed RMS roughness as low as that of the substrate (~1 nm) indicating extremely smooth surfaces. The combination of the presented AFM images and Scanning Electron Microscopy (SEM) images (see SI4) allows discarding a nano-platelet structure as reported previously in similar metal-organic films.[17]

We also employed the Langmuir–Schäfer technique to deposit the films on regular copper TEM grids (Figure 3a) and on $SiO_2$ substrates (Figure 3b) with predefined circular wells with diameters ranging from 0.5 to 3 µm. Surprisingly we found that during the Langmuir–Schäfer process the films were suspended over these holes of $SiO_2$ substrates not collapsing to the substrate. AFM images of these micro-drums are shown in Figure 3b and 3c. The obtaining of these free-standing films is already indicative of mechanical robustness since most of 2D materials studied so far tend to collapse or break during transfer to this type of substrates due to capillary forces in wet processes.[27] These nano-drums allowed us to measure their mechanical properties by indentation experiments with AFM tips. Indentation curves were performed with Si tips of ~2.8 N/m spring constant and 20 nm radius. Under these conditions, the radius of the tip is much smaller than the radius of the drumhead and therefore the Force *vs.* Indentation *F(δ)* curve in the elastic region can be well approximated to:[28]

$$F(\delta) = \pi T\delta + \frac{Etq^3}{a^2}\delta^3 \qquad \text{Eq. 1}$$

where F is the loading force, δ is the indentation at the central point, T is the pretension accumulated in the sheet during the preparation procedure, q ≈ 1 is a factor that accounts for the *Poisson* ratio of the material, $a$ is the drumhead radius, t is the thickness of the measured membrane and E is the elastic modulus of the film. A representative *F(δ)*, performed on the drumhead shown in Figure 3c, is depicted in Figure 3d. The fitting of our experimental curves to Eq. 1 yielded values of E = 11 ± 3 GPa.

In addition to the stiffness, we could also observe the yield point of the membranes by loading some drumheads up to the failure point that was at 170 nN with a 30 nm tip radius. The yield strength can be roughly estimated[29] by $\sigma = \sqrt{F_{max}E/(t\pi R_{tip})}$ , where $F_{max}$ is the force at which fracture takes place and $R_{tip}$ is the radius of the AFM tip. This leads to an estimation of $\sigma$=1.0 ± 0.3 GPa, showing that these ultrathin films can sustain quite large deformations without breaking (notice that this simple expression for the breaking force is for a linear material and tends to overestimate this figure). For the sake of comparison, the commonly accepted values for E and $\sigma$ in graphene are 1 TPa and 0.14 TPa respectively.[30] Importantly, even for loads higher than the yield points we have never observed catastrophic failure of our films (characteristic of laminar inorganic crystals), in fact the notch created by the tip does not propagate through the films. This observation suggests the presence of defects in the atomic lattice that stabilize the films against mechanical failure.[27] Besides the concrete values of these mechanical properties, that are lower to those reported in covalent polymers,[31] our observations sustain that, contrary to what one might expect, the weak coordination bonds allow producing robust, ultrathin, flexible and mechanically stable films.

Electrical characterization of $[Cu_2I_2(TAA)]_n$ films was performed by providing gold electrodes on top of the films by thermal evaporation using stencil masks (Figure 2f and Figure 4a). The highly doped silicon substrate underneath the insulating $SiO_2$ was eventually used to apply a gate voltage. Figure 4b depicts typical current *vs.* bias voltage curves (IV curves), acquired at high (> 1 V/s, blue line) and low (< 0.3 V/s, red line) bias sweeping speeds. A initial electrical characterization of $[Cu_2I_2(TAA)]_n$ films was performed at fast bias sweepings leading to linear IV curves, indicating ohmic behaviour. However, this initial characterization also shows that conductivity of the films spanned over a wide range, from $10^{-4}$ S/cm (a typical value for an organic conductor[16]) to 50 S/cm (comparable to that of best conjugated polymers as polypyrrole[32]) depending on the region measured, with no dependence on device channel length -from 2 to 30 µm- or film thickness -from 4 to 60 nm-.

For sweeping speeds below 0.3 V/s most of our devices showed pinched-hysteresis loops in IV curves at room temperature. The size of the loops increases as the voltage sweep rate is reduced. These hysteresis loops showed two resistance states with $R_{OFF}/R_{ON}$ ratio up to 30. A representative IV curve is depicted in red in Figure 4b. These features in the IV curves are typical of memristive or resistive-switching systems, where the resistance depends on the history of the current previously flowed through the device. The mechanism behind this response will be discussed below.

Three terminal measurements with a global back gate showed very weak dependence of current upon an external electric field (Figure 4c). We also investigated the impact of metal contacts with different work function materials but did not observe any relevant change in the conductivity values. Neither had atmosphere an influence on the electrical properties: equivalent measurements in ambient conditions (T = 25 °C, relative humidity 55 %) and high vacuum did not show any difference in conductivity. Interestingly, high conductivity films reached current densities of $2\times10^{-4}$ ?A/$nm^2$ at room temperature. Out-of-plane conductivity was evaluated by collecting films on gold surfaces and contacting them with $Ga_2O_3$/EGaIn as top electrode (see SI5). The obtained values of $10^{-8}$ S/cm evidence a strong anisotropic conduction in the films.

Upon cooling from 300 to 70 K the films showed an in-plane conductivity decrease of one order of magnitude. No single conduction law can fit the entire curve, indicating that there are at least two conduction mechanisms contributing to the conductivity in different temperature ranges as observed in other polymers.[16] As depicted in Figure 4d, the threshold temperature is around 210 K. Arrhenius plots show that, while the characteristic activation energy for electrical conduction at low temperatures is 12 ± 1 meV, it increases up to 170 ± 20 meV for temperatures above 200 K. Furthermore, AC conductivity measurements did not show any dependence from 1 to $10^7$ Hz. The low activation energies together with the AC measurements and the insensitivity of electrical transport to air exposure strongly suggest electronic rather than ionic (proton) conductivity as reported in similar polymers.[33]

Further hint from the physical process underlying electrical conduction in our films came from observations in single-crystals of $[Cu_2I_2(TAA)]_n$. Amorphous crystals showed much higher conductivity than single crystals, which prompted us to look for signatures of amorphization in our films. The planar nature of our devices and their in-plane conductivity (SI5) provide an excellent platform for the *in situ* acquisition of local surface potential maps by Kelvin Probe Force Microscopy (KPFM). Figure 5 depicts representative data of high conductivity devices ($\sigma > 1$ S/cm). Figures 5a and 5b show the topography and the simultaneously acquired surface potential map of a non-operating device. Here we observe that the bottom right area of the film presents an assembly of patches several μm in lateral size with surface potential differences of few hundred millivolts, in contrast with the featureless top left region. Surface potential images of this device recorded *in-operando,* shown in Figure 5c, reveal that in the regions with no observable patches in panel b, the voltage drop takes place mainly at the contacts with the metal electrodes (red profile in Figure 5c). However, in the regions where we observed patches, the voltage drop occurs all through the film length (green profile in Figure 5c). This ensemble of images allowed establishing a direct correlation between the areas rich in surface potential patches and high conducting regions. As expected from these images, devices with low conductivity ($\sigma$ < 0.1 S/cm) did not show any features in the KPFM images (SI6). This, along with observations in crystals, make plausible to ascribe the observed patches/regions in surface potential to different crystallographic structures/orientations determined by X-ray diffraction (as it has been previously observed in other 2D materials[34, 35]). It thus follows that the mechanism governing charge transport in our films is grain boundary mediated. In common conductive (ohmic) materials the presence of grain boundaries usually leads to a decrease in conductivity, however accumulation of certain charge carriers and appearance of mid-gap states at these domain walls in insulating materials might render the regions near boundaries much more conductive than the bulk and provide paths for charge movement as reported in related coordination polymers crystals.[33] While the structural disorder revealed by X-ray spectroscopy makes it difficult to define the atomic structure at these interphases, accumulation of iodine or sulfur vacancies, as already observed in $MoS_2$,[35] are good candidates to explain such effect. Independently of the specific cause, grain boundary conduction is certainly a plausible cause to account for the high dispersion of values found for the conductivity of our ultrathin films: areas rich in boundaries would show much high conductivity.

As mentioned above, our devices also showed memristive behaviour at low frequency bias sweeps. Surface potential maps of memristive devices with high applied bias (> 3 V) revealed boundary migration (see SI7), suggesting that this is the main mechanism involved in memristive behaviour. Further support to the proposed boundary mediated conduction comes from the observation that the slow initial sweeps in bias of our films are typical of space charge limited current (SCLC) (see SI8). This conduction mechanism is characteristic of dielectric solids with trapped charges due to spatially inhomogeneous resistance:[36] IV curves show a power-law dependence $I \propto V^n$ where $n$ increases with bias: in Figure S11b we can distinguish an ohmic part at low voltages, a second region with $I \propto V^2$ where the traps begin to be filled with the injected carriers and a third region with $n > 2$ where all traps are filled up, so the subsequently injected carriers can move in the dielectric film.

In summary, we report a bottom-up approach to fabricate CP films, based on cheap building blocks (CuI and TAA), with down to 4 nm thickness extended across macroscopic regions. Interestingly, the films withstand free-standing geometry with elastic constants of 11 GPa and 1 GPa for the Young's modulus and yield strength, respectively. Additionally, these CP films exhibit in-plane electrical conductivity up to 50 S/cm that is controlled by the presence of grain boundaries. The easy creation and motion of these boundaries suggest that they could be manipulated to create novel devices with tuneable electro-mechanical properties. With the aim of enhancing the performance of these films, current work is focused on the controlled creation of grain boundaries by temperature quenching methods.

**FIGURES**

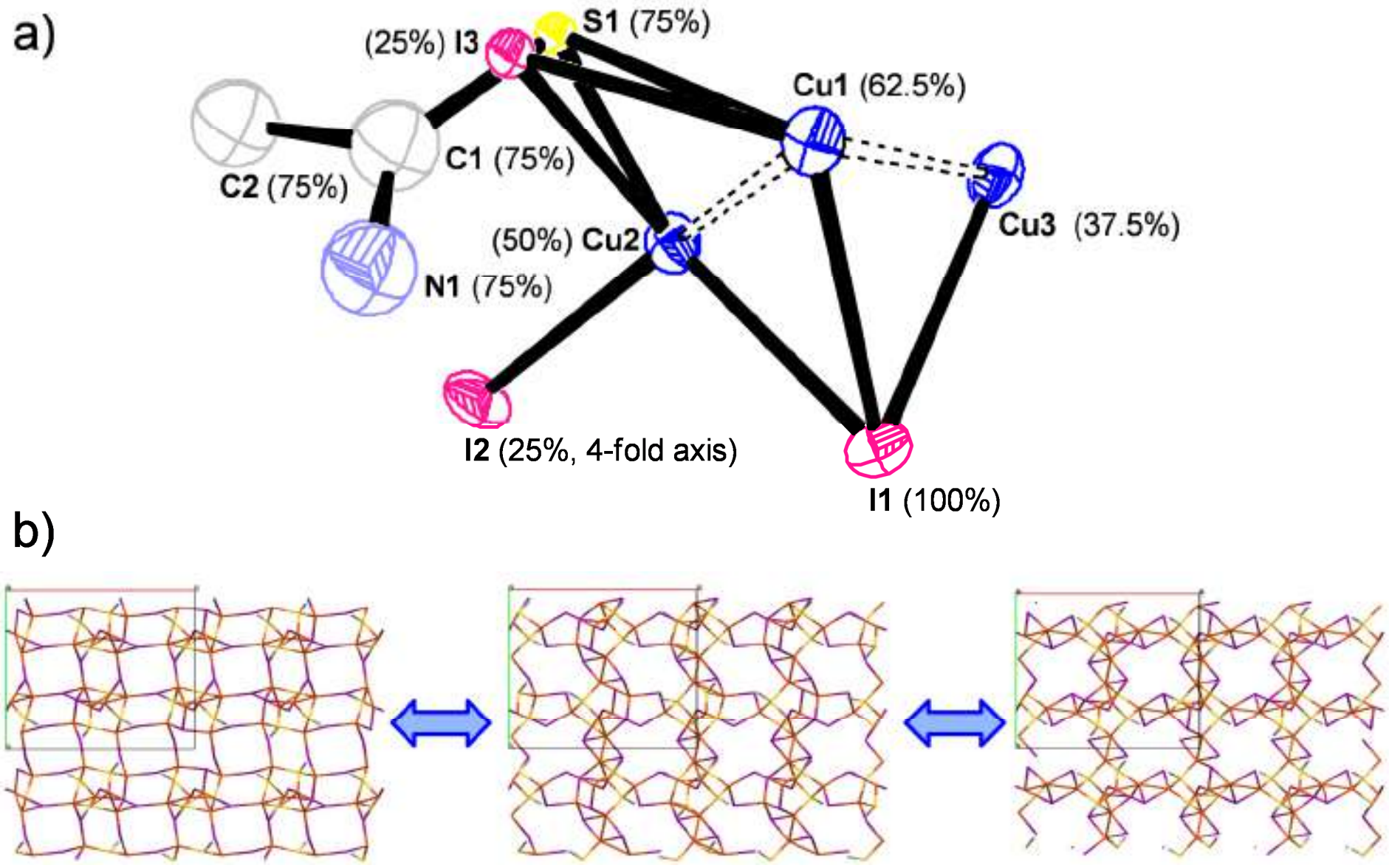

**Figure 1. Structure of $[Cu_2I_2(TAA)]_n$ compound.** a) ORTEP (Oak Ridge Thermal-Ellipsoid Plot Program) view of the asymmetric unit of compound $[Cu_2I_2(TAA)]_n$ emphasizing the occupation factors. The disorder of the thioacetamide ligand was omitted for sake of clarity. b) Some plausible ordered models contributing to the highly disordered final crystal architecture obtained from single crystal X-ray diffraction.

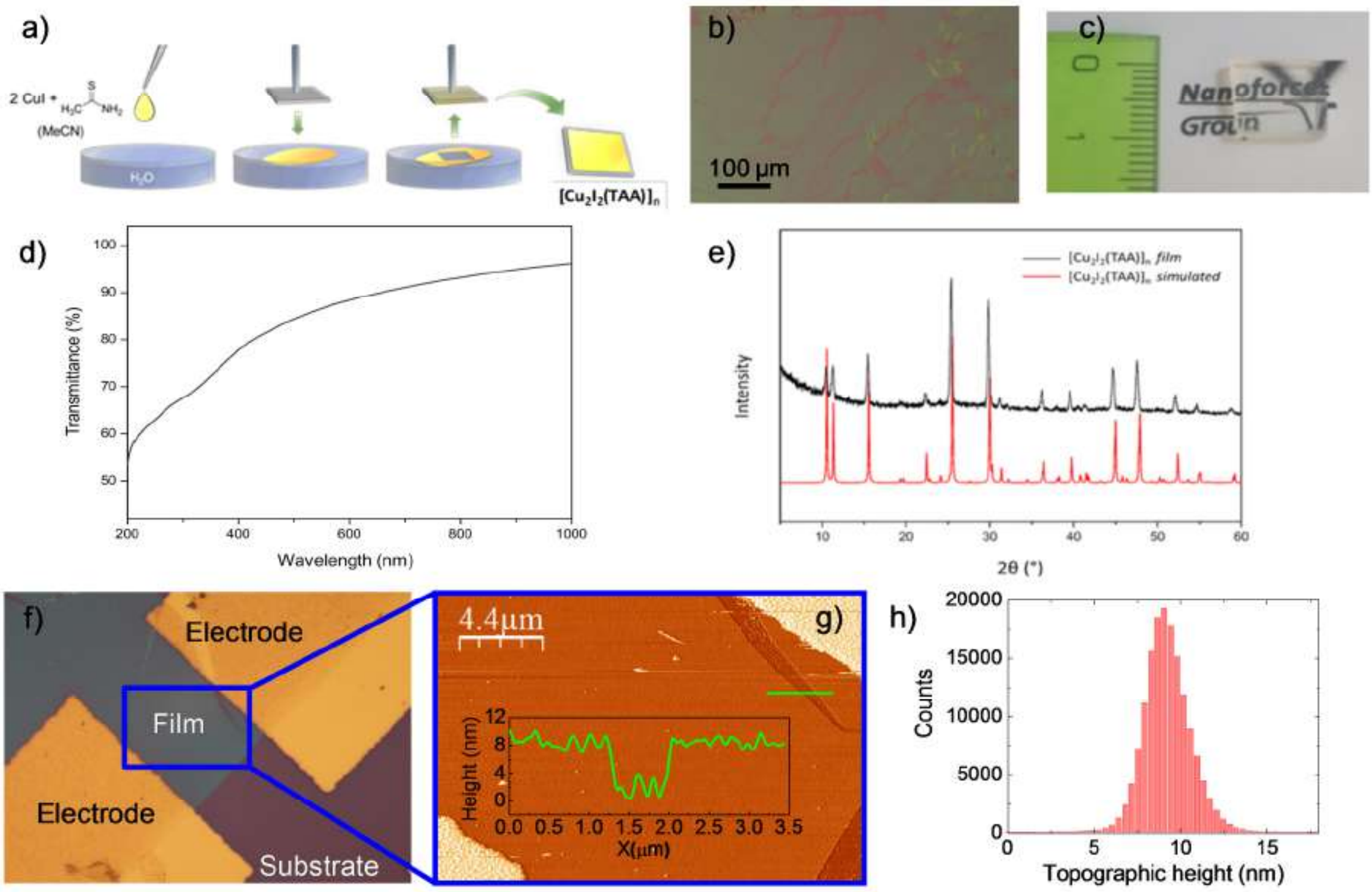

**Figure 2. Preparation and characterization of $[Cu_2I_2(TAA)]_n$ films.** a) Scheme of the $[Cu_2I_2(TAA)]_n$ film preparation at the water-air interface and the Langmuir–Schäfer technique used to collect it on surfaces. b) Optical image of a representative sample deposited on a $SiO_2$ substrate. c) Photograph of a $[Cu_2I_2(TAA)]_n$ film collected on fused quartz with a ruler as scale bar. d) UV-visible spectrum of the film in c). e) GI X-ray diffraction (incidence angle, α = 0.5°) of the $[Cu_2I_2(TAA)]_n$ film collected on $SiO_2$ and its comparison with that calculated from the single crystal X-ray diffraction data (red). f) Optical image of a film electrically contacted by two gold electrodes. g) AFM topographic image of the area inside the blue rectangle marked in panel f). The inset is a profile corresponding to the green horizontal line along the film and a crack in it. From this profile a film thickness of 8 nm is obtained. h) Histogram of topographic heights obtained on a 10 × 10 $\mu m^2$ area in the film shown in panel g). From this histogram RMS roughness of 1.2 nm is obtained.

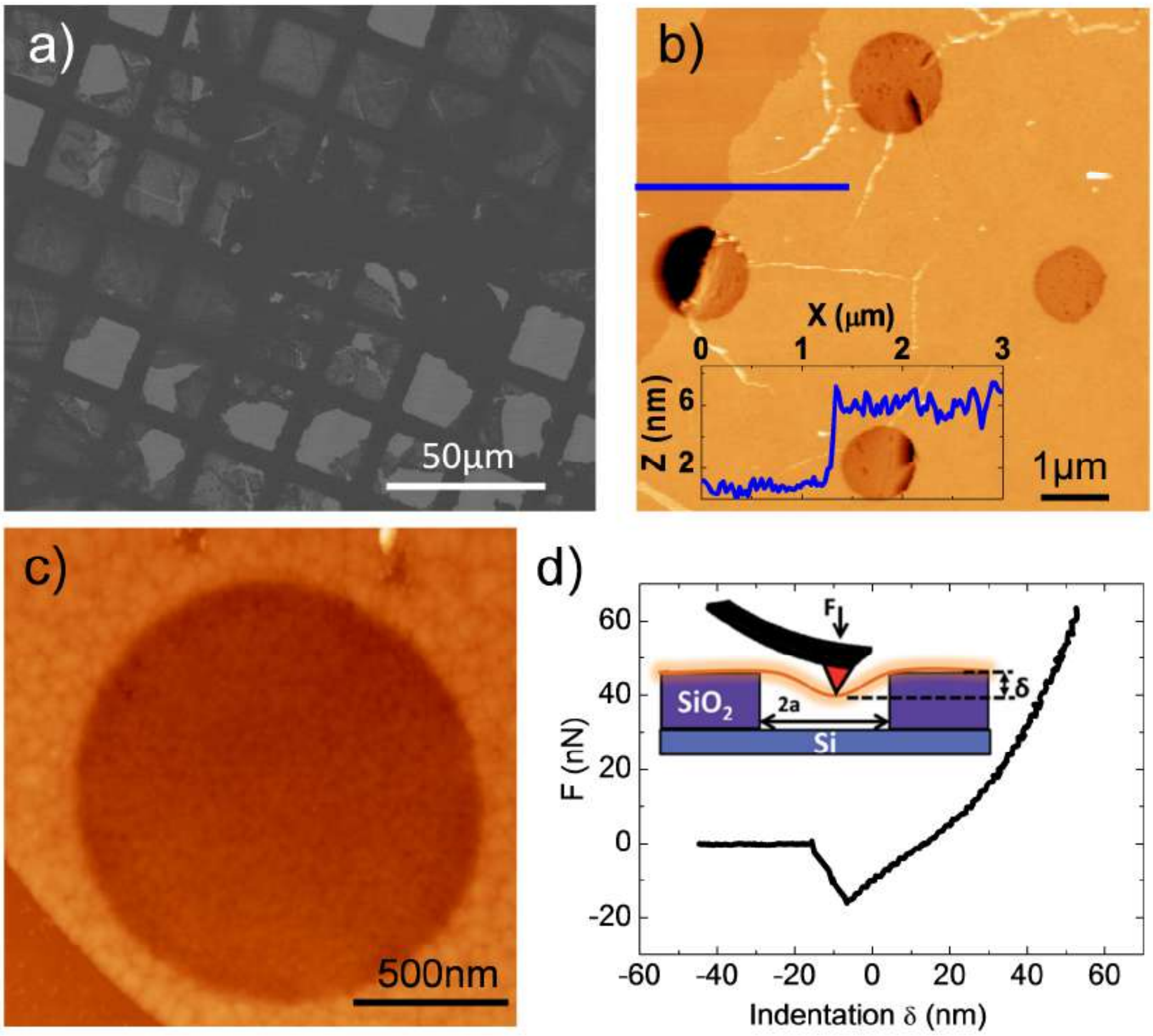

**Figure 3. Mechanical robustness of $[Cu_2I_2(TAA)]_n$ films.** a) TEM image of a $[Cu_2I_2(TAA)]_n$ film on a regular copper TEM grid. Dark areas correspond to unsupported $[Cu_2I_2(TAA)]_n$ films. b) Representative AFM image of a 6 nm thick film suspended over several circular wells. Here we observe that even those regions presenting cracks remain suspended. The inset is the height profile along the blue line in the image. c) Representative AFM image of a 1.5 µm diameter $[Cu_2I_2(TAA)]_n$ drum where indentations were performed. d) Representative F(δ) curve on a 10 nm thick drum. The inset shows a lateral view of the experimental set up used to perform nanoindentations.

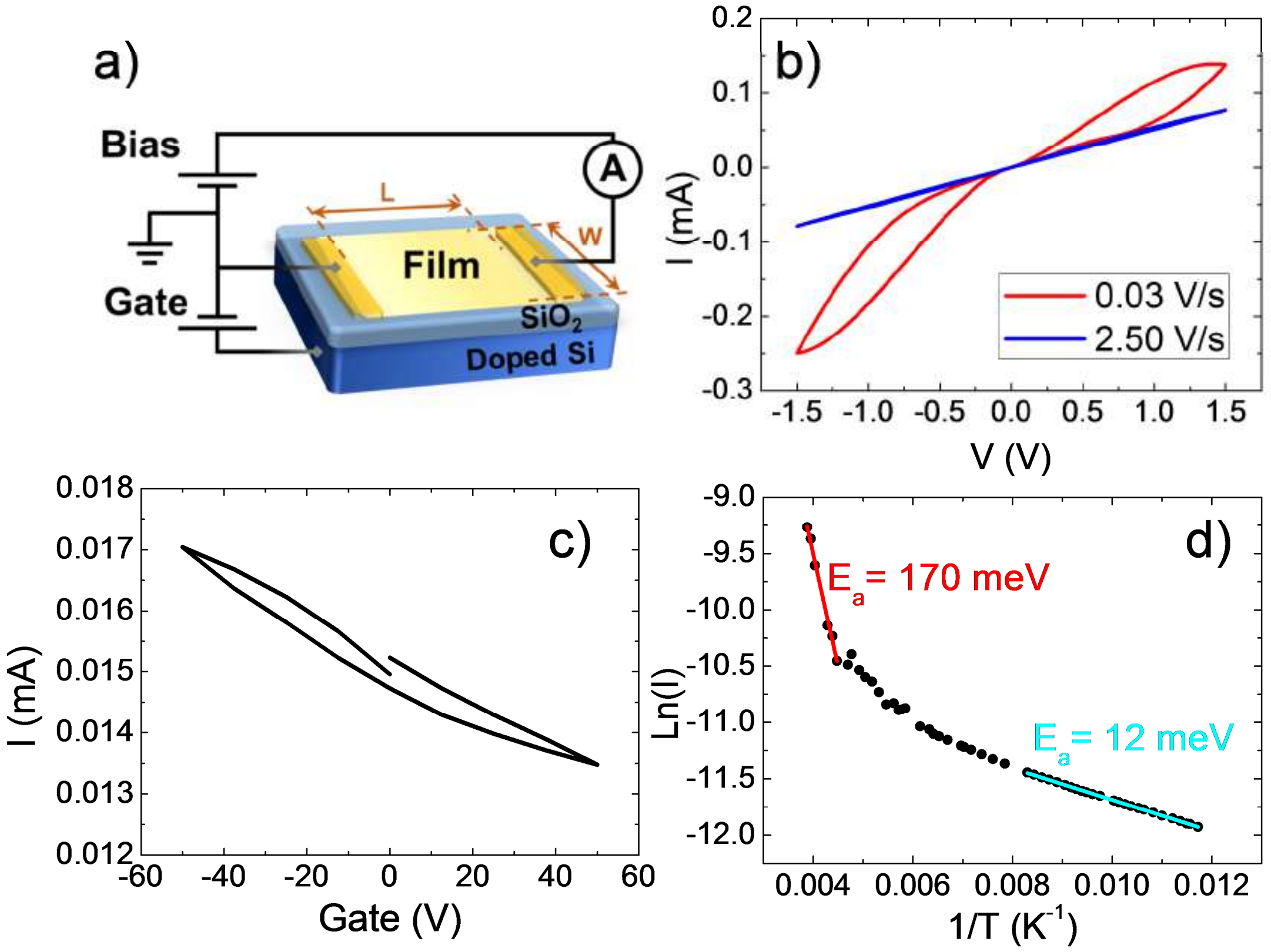

**Figure 4. Electrical properties of $[Cu_2I_2(TAA)]_n$ films.** a) Scheme of the set up used to electrically characterize the films. b) Two representative IV curves acquired on the same $[Cu_2I_2(TAA)]_n$ film at two bias sweeping speeds, where metallic and memristive behaviours are shown. c) Current *vs.* gate voltage at a fixed bias of 3 V. d) Logarithm of the current *vs.* inverse temperature for a high conductivity device, lines are linear fittings to the corresponding range of data.

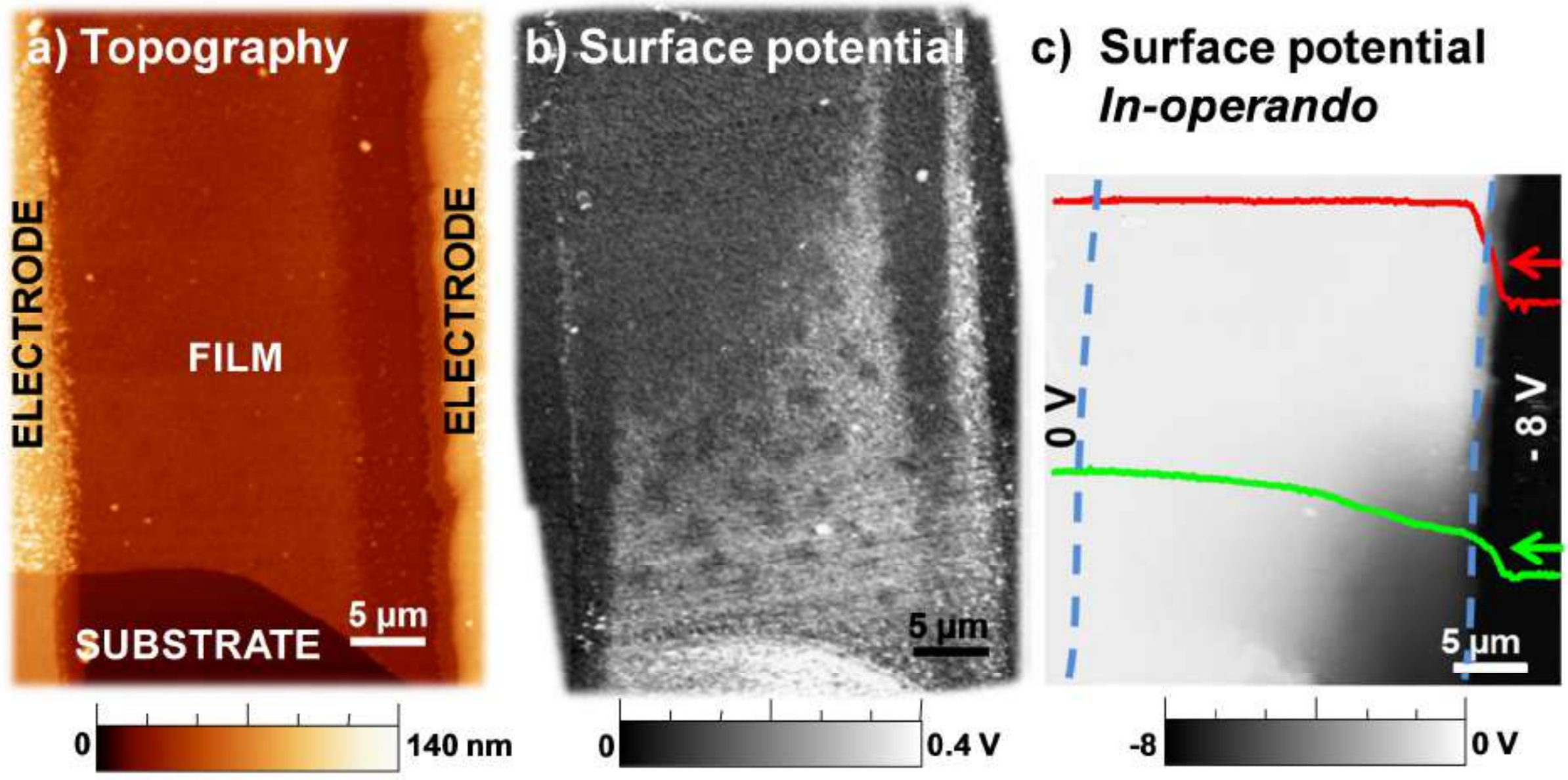

**Figure 5. Surface potential maps of $[Cu_2I_2(TAA)]_n$ films.** a) AFM topographic image of a representative device with no significant topographic features. b) Surface potential map acquired by KPFM in the same area as panel a) without applying an electrical field between the electrodes. c) Surface potential map acquired in the same area as the previous panels while -8 V were applied to the right electrode. The red and green lines are the surface potential profiles across the lines indicated by the arrows in corresponding colours. The blue dashed lines are drawn at the electrode borders.

## Methods

**$[Cu_2I_2(TAA)]_n$ Film Preparation.** Nanosheets were prepared by addition of 50 µL of a solution containing the building blocks ([CuI] = 0.1 M; [thioacetamide] = 0.05 M) in acetonitrile at the air/water interface, using Milli-Q water contained in a Petri dish (ø = 4.5 cm). The formation of nanosheets takes place immediately. Films were deposited by horizontal lift onto the substrate by Langmuir–Schäfer (LS) method. The substrate was dried with a flow of Ar or $N_2$ gas.

**Bulk Synthesis of $[Cu_2I_2(TAA)]_n$.** All reactions were carried out at 20 °C and under ambient conditions. A solution containing CuI (0.097 g, 0.50 mmol) and thioacetamide (0.019 g, 0.25 mmol) in acetonitrile (5 mL) was slowly poured, dropwise, into 100 mL of water contained in a beaker. The resulting suspension was then collected by filtration, and the yellow solid was washed with water and diethyl ether. Yield: 0.078 g (71 %). Anal. Calcd (%) for $C_2H_5Cu_2I_2NS$: C, 5.27; H, 1.11; N, 3.07; S, 7.03. Found: C, 4.93; H, 1.05; N, 2.76; S, 6.94. Slow diffusion at 25 °C diethyl ether into a 0.3 $molL^{-1}$ solution of CuI and thioacetamide, 2:1 ratio, in acetonitrile (5 mL) gives rise to single crystals suitable for X-ray diffraction.

**Instrumentation.** The Atomic Force Microscope (AFM) morphological characterization was performed in dynamic mode using commercial Nanosensors cantilevers of 3 and 40 N/m nominal stiffnesses, with resonance frequencies of about 75 and 300 kHz, respectively. For KPFM experiments, commercial Budget Sensors Cr/Pt conductive cantilevers with 3 N/m nominal force constant and 75 kHz nominal resonance frequency were used. All AFM data were acquired with a modified Nanotec Electrónica microscope controlled with the SPM software package WSxM,[37] which was also used to process them.

The fabrication of electrodes for the electrical characterization of the samples was carried out using thermal gold evaporation with a stencil mask after film deposition. Specifically, 100 nm of gold were deposited on the samples at a very low deposition rate (0.1 Å/s) under high-vacuum ($10^{-6}$ mbar). We electrically characterized our devices in a home-made three terminal probe station equipped with a Keithley 2400 sourcemeter, a Keithley 2000 multimeter and a variable-gain low noise current amplifier. It allowed variable temperature measurements in the range 70 – 400 K. The probe station was located inside a high-vacuum chamber (base pressure $1\times10^{-6}$ mbar) to avoid water condensation when measuring at low temperatures.

The X-ray diffraction data collections and structure determinations were done at 100, 200 and 296 K on a Bruker Kappa Apex II diffractometer using graphite-monochromated Mo-Kα radiation (λ = 0.71073 Å). The cell parameters were determined and refined by a least-squares fit of all reflections. A semi-empirical absorption correction (SADABS) was applied for all cases. All the structures were solved by direct methods using the SIR92 program[38] and refined by full-matrix least-squares on $F^2$ including all reflections (SHELXL97).[39] All calculations were performed using the WINGX crystallographic software package. Relevant data acquisition and refinement parameters are gathered in SI1 in Table S1.

The films were characterised by grazing incident X-ray diffraction (GIXRD) with a Panalytical X'Pert PRO in the range of 2θ = 3 – 50°, with an increment of 0.02°, under an incidence angle of 0.5°.

We conducted the X-ray photoelectron spectroscopy (XPS) at the SINS (Surface, Interface and Nanostructure Science) beamline of Singapore Synchrotron Light Source (SSLS) with a base pressure of $1\times10^{-10}$ mbar. A sputter-cleaned gold foil in electrical contact with the sample was used to calibrate the photon energy (PE) with the 84.0 eV of Au 4f7/2 core level peak. Here, we chose the PE at 350 eV to probe the C 1s, S 2p and Au 4f, at 650 eV to probe the O 1s and N 1s, 800 eV to probe the I 3d, and 1100 eV to probe the Cu 2p. For the Cu 2p spectra, we applied -10.0 V or +10.0 V to the sample to monitor any changes of spectra under bias. Figure S4 shows all the high resolution XPS spectra.

**Supporting Information**. A CIF file with the crystallographic information and a pdf file containing additional structural and electrical characterization data is available online.

**Corresponding Authors**

*E-mails : cristina.gomez@uam.es; felix.zamora@uam.es

**Author Contributions**

The manuscript was written through contributions of all authors. All authors have given approval to the final version of the manuscript.

**Acknowledgments.** We acknowledge financial support through the "María de Maeztu" Programme for Units of Excellence in R&D (MDM-2014-0377) and from projects MAT2016-77608-C3-1-P and 3-P, MAD2D-CM and MAT2013-46753-C2-2-P that includes Miriam Moreno-Moreno's FPI fellowship. Ramon Areces foundation is acknowledged for financial support. We acknowledge the Ministry of Education (MOE) for supporting this research under award No. MOE2015-T2-2-134 Prime Minister's Office, Singapore under its Medium sized centre program is also acknowledged for supporting this research. The authors would like to acknowledge the Singapore Synchrotron Light Source (SSLS) for providing the facilities at the Surface, Interface and Nanostructure Science (SINS) beam line under NUS core support C-380-003-003-001. The Laboratory is a National Research Infrastructure under the National Research Foundation Singapore.

# SUPPORTING INFORMATION

# One-Pot Preparation of Mechanically Robust, Transparent, Highly Conductive and Memristive Metal-Organic Ultrathin Film

Miriam Moreno-Moreno, Javier Troyano, Pablo Ares, Oscar Castillo, Christian A. Nijhuis, Li Yuan, Pilar Amo Ochoa, Salomé Delgado, Julio Gómez-Herrero, Félix Zamora and Cristina Gómez-Navarro

## SI1. X-ray structure of $[Cu_2I_2(TAA)]_n$

**X-ray diffraction data collections and structure determinations.** They were done on a Bruker Kappa Apex II diffractometer using graphite-monochromated Mo-Kα radiation (λ = 0.71073 Å). The cell parameters were determined and refined by a least-squares fit of all reflections. A semi-empirical absorption correction (SADABS) was applied for all cases. All the structures were solved by direct methods using the SIR92 program and refined by full-matrix least-squares on $F^2$ including all reflections (SHELXL97). All calculations were performed using the WINGX crystallographic software package.[1] To solve the crystal structure of this compound it was necessary to introduce a great disorder on the position and occupation of the copper(I) metal centres and the thioacetamide ligand. The latter was refined imposing geometrical restraints on the planarity and C-C and C-N distances. The hydrogen atoms of these highly disordered thioacetamide ligands were not located. The iodide anions are the only atoms that remain non-disordered in the crystal structure. All the attempts to be able to reproduce the experimental data using lower symmetry space groups did not remove the presence of this disorder. All non-hydrogen atoms except those belonging to the thioacetamide ligand were refined anisotropically. Relevant data acquisition and refinement parameters are gathered in Table S1. CCDC 1568566-1568568 contain the supplementary crystallographic data for this paper.

These crystallographic features imply that the disorder cannot be explained as a mixture of two ordered models. In this compound, due to the great extent of the disorder, the amount of possible ordered models contributing to the mixture is huge and it has become impossible to provide a detail on all these ordered models. However, we provide some insight into three of these possible ordered models in which we can observe how the crystal structure is always build up by means of tetrahedral copper(I) metal centres. In all cases, the copper-iodide and copper-sulphur distances lay within their usual values.

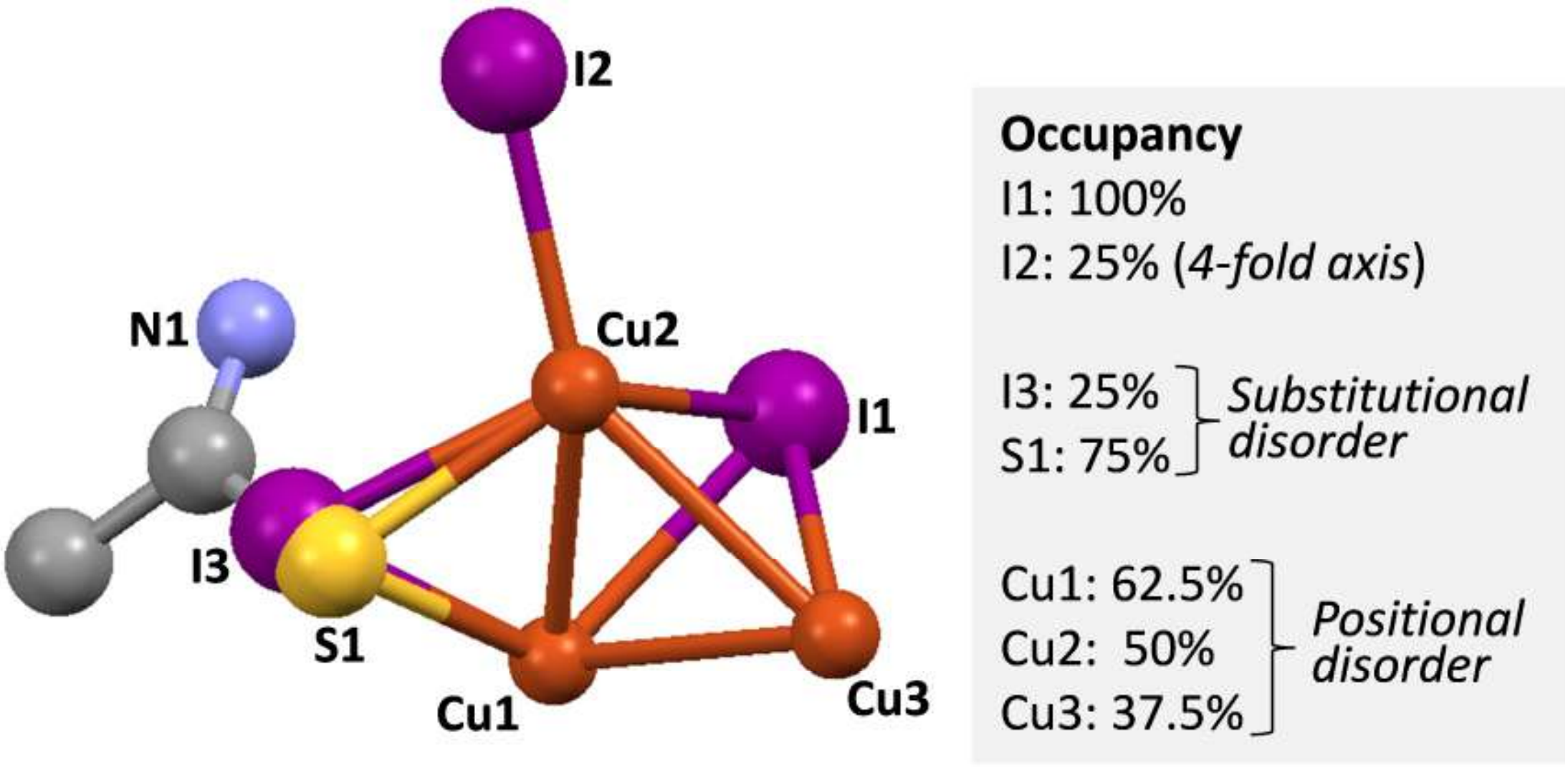

**Figure S1.** View of the asymmetric unit of $[Cu_2I_2(TAA)]_n$ compound showing the occupation factors.

# *A* MODEL

There are seven non-equivalent metal centres, all of them involve a tetrahedral coordination environment (with donor sets ranging from $I_4$, $I_3S$ to $I_2S_2$). It must be emphasized also that for clarity purposes the non-sulphur atoms of the thioacetamide ligand have been omitted.

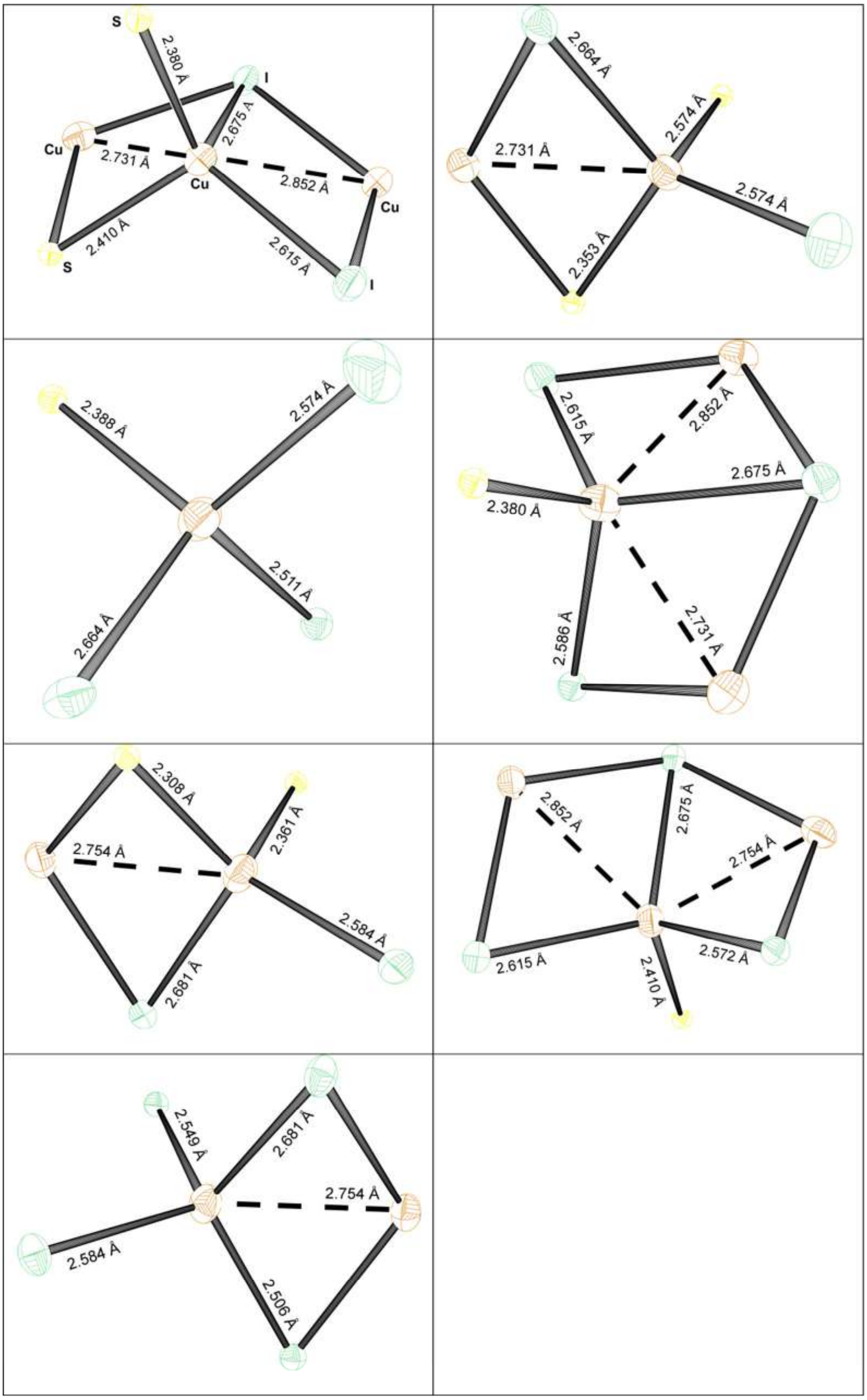

# *B* MODEL

There are seven non-equivalent metal centres, all of them involve a tetrahedral coordination environment (with donor sets ranging from $I_3S$ to $I_2S_2$). It must be emphasized also that for clarity purposes the non-sulphur atoms of the thioacetamide ligand have been omitted.

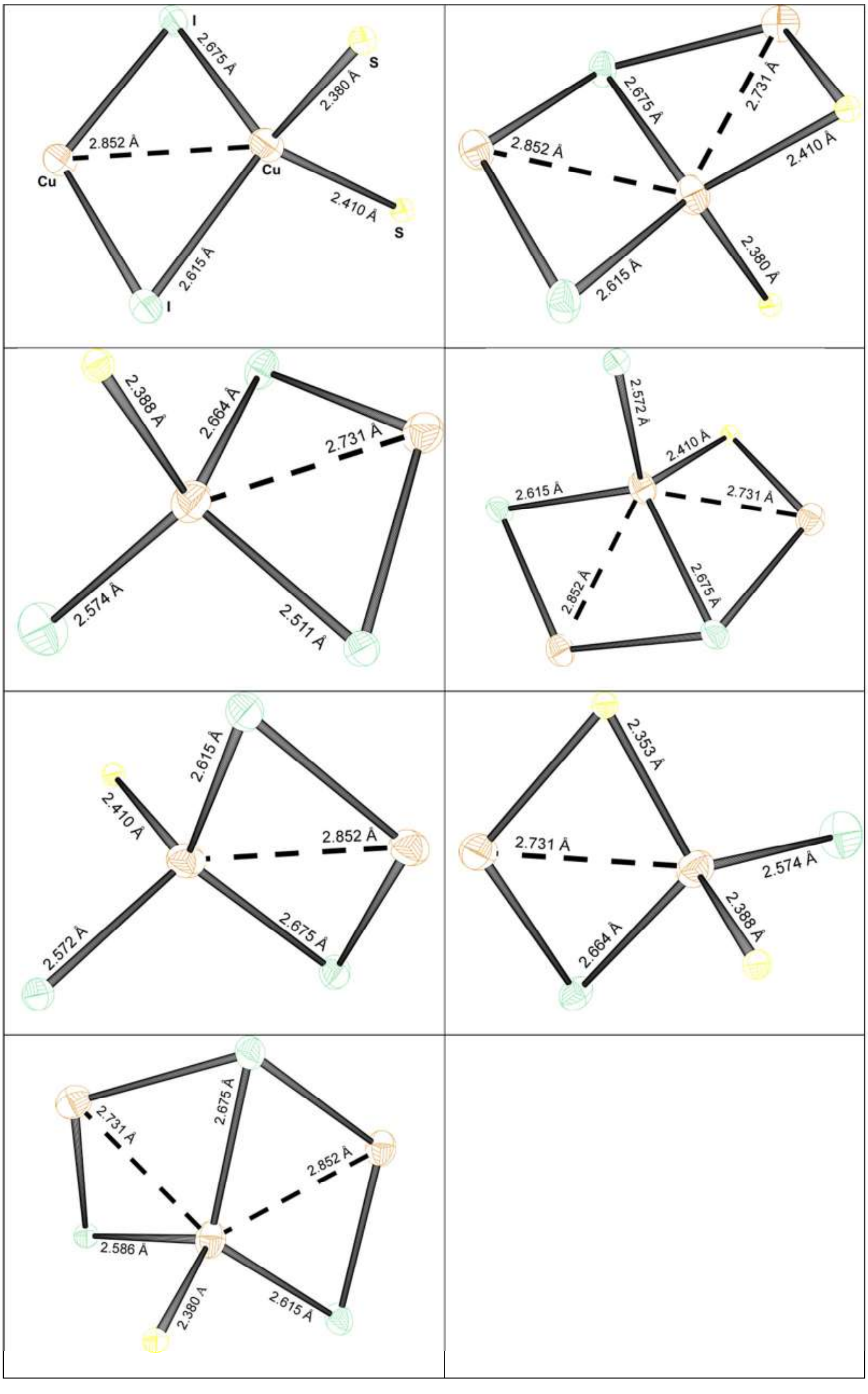

# *C* MODEL

There are nine non-equivalent metal centres, all of them involve a tetrahedral coordination environment (with donor sets ranging from $I_3S$ to $I_2S_2$). It must be emphasized also that for clarity purposes the non-sulphur atoms of the thioacetamide ligand have been omitted.

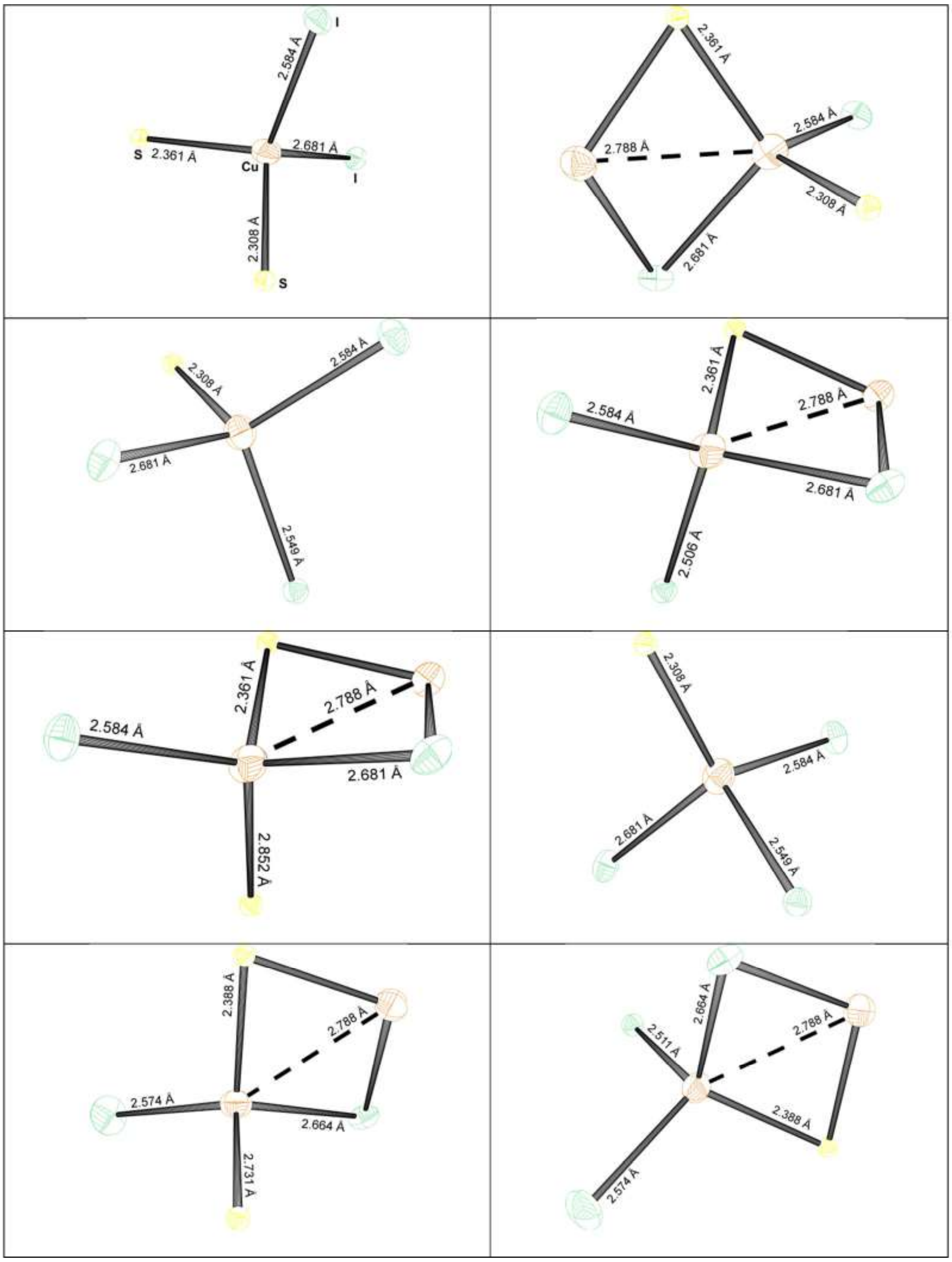

**Table S1.** Crystallographic data for $[Cu_2I_2(TAA)]_n$ compound at different temperatures.

| | $[Cu_2I_2(TAA)]_{n\mathbf{100K}}$ | $[Cu_2I_2(TAA)]_{n\mathbf{200K}}$ | $[Cu_2I_2(TAA)]_{n\mathbf{296K}}$ |
|---|---|---|---|
| Empirical formula | $C_2H_5Cu_2I_2NS$ | $C_2H_5Cu_2I_2NS$ | $C_2H_5Cu_2I_2NS$ |
| Formula weight | 456.01 | 456.01 | 456.01 |
| T (K) | 100(2) | 200(2) | 296(2) |
| Crystal system | Tetragonal | Tetragonal | Tetragonal |
| Space group | $I\,4_1/a$ | $I\,4_1/a$ | $I\,4_1/a$ |
| $a$ (Å) | 15.5848(7) | 15.6261(18) | 15.6634(3) |
| $c$ (Å) | 9.8950(5) | 9.9155(13) | 9.9466(2) |
| V (Å$^3$) | 2403.4(2) | 2421.1(6) | 2440.32(11) |
| Z | 12 | 12 | 12 |
| $R_{int}$ | 0.1278 | 0.1241 | 0.1233 |
| Reflections collected | 10746 | 18868 | 19065 |
| Independent reflections | 1107 | 1245 | 1251 |
| Reflections [I>2σ(I)] | 976 | 1047 | 1014 |
| Parameters | 72 | 72 | 72 |
| Restraints | 9 | 9 | 9 |
| Goodness of fit (S)[a] | 1.077 | 1.076 | 1.065 |
| R1[b] [I>2σ(I)] | 0.0389 | 0.0488 | 0.0530 |
| wR2[c] [I>2σ(I)] | 0.1065 | 0.1293 | 0.1427 |
| R1[b] [all data] | 0.0438 | 0.0550 | 0.0609 |
| wR2[c] [all data] | 0.1091 | 0.1341 | 0.1484 |
| Largest peak/hole ($e^-$ Å$^3$) | 1.007/-1.780 | 2.337/-2.211 | 2.262/-2.001 |

[a] $S = [\sum w(F_0^2 - F_c^2)^2 / (N_{obs} - N_{param})]^{1/2}$ [b] $R1 = \sum||F_0|-|F_c|| / \sum|F_0|$; [c] $wR2 = [\sum w(F_0^2 - F_c^2)^2 / \sum wF_0^2]^{1/2}$; $w = 1/[\sigma^2(F_0^2) + (aP)^2 + b]$ where $P = (max(F_0^2,0) + 2\,F_c^2)/3$ with a = 0.0554 ($[Cu_2I_2(TAA)]_{n\ \mathbf{100K}}$), 0.0903 ($[Cu_2I_2(TAA)]_{n\ \mathbf{200K}}$), 0.1015 ($[Cu_2I_2(TAA)]_{n\ \mathbf{296K}}$) and b = 24.6784 ($[Cu_2I_2(TAA)]_{n\ \mathbf{100K}}$), 2.2898 ($[Cu_2I_2(TAA)]_{n\ \mathbf{200K}}$).

## SI2. Electrical conductivity of bulk crystals

The electrical conductivity of crystals was measured by contacting them with graphite paint. Electrical characterization of obtained crystals showed that good quality crystals (selected by optical inspection of their morphology) presented conductivity values of $10^{-5}$ - $10^{-3}$ S/cm, but crystals with irregular morphology presented values up to 10 S/cm. To corroborate this suggested relation between crystallinity and electrical conductivity we tested good quality crystals (as indicated by X ray diffraction) and annealed them with the aim of amorphizating the structure, measuring electrical conductivity on the same crystal before and after annealing.

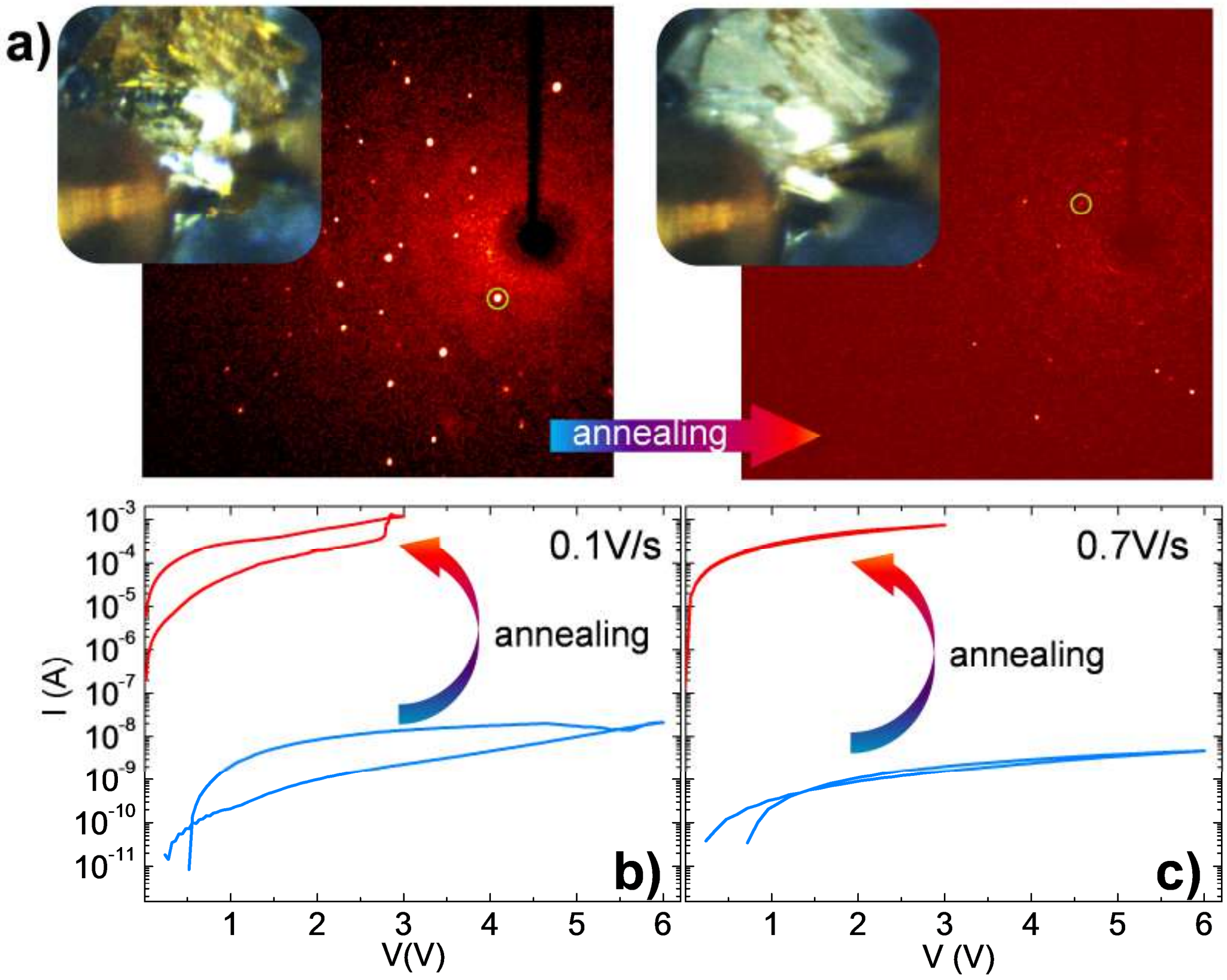

**Figure S2**. a) Intense and well defined diffraction spots are observed for a single crystal that has been kept under room conditions (left), but significantly less intense diffraction spots are found in the same crystal upon heating up to 70 °C (right), indicative of a loss of crystallinity. The (420) diffraction peak, in the centered tetragonal system *I*, has been encircled in both images for comparative purposes. In both cases the same exposure time was employed. The insets show optical images of the $[Cu_2I_2(TAA)]_n$ crystal before and after annealing from RT to 70 °C. The amorphization can be already appreciated in the brightness of the crystals. Panels b) and c) show *IV* curves on the crystal in a semi-log plot before (blue) and after (red) annealing for two sweep rates. The crystals showed an increase in conductivity of 5 orders of magnitude upon annealing and showed similar memristive character as the films.

## SI3. XPS, IR spectra and thermal stability

**Table S2.** Binding energy values for different $[Cu_2I_2(TAA)]_n$ samples determined from XPS measurements.

| Sample | Binding energy (eV) | | | | |
|---|---|---|---|---|---|
| | $Cu2p_{3/2}$ | $I3d_{5/2}$ | S2p | N1s | C1s |
| $[Cu_2I_2(TAA)]_n$ powder | 932.5 | 619.4 | 162.8 | 399.9 | 284.8 (65)<br>286.2 (35) |
| $[Cu_2I_2(TAA)]_n$ crystal | 932.6 | 619.3 | 162.9 | 399.9 | 284.8 (67)<br>286.3 (33) |
| $[Cu_2I_2(TAA)]_n$ film | 932.5 | 619.4 | 162.9 | 399.3 | 284.8 (65)<br>286.3 (35) |

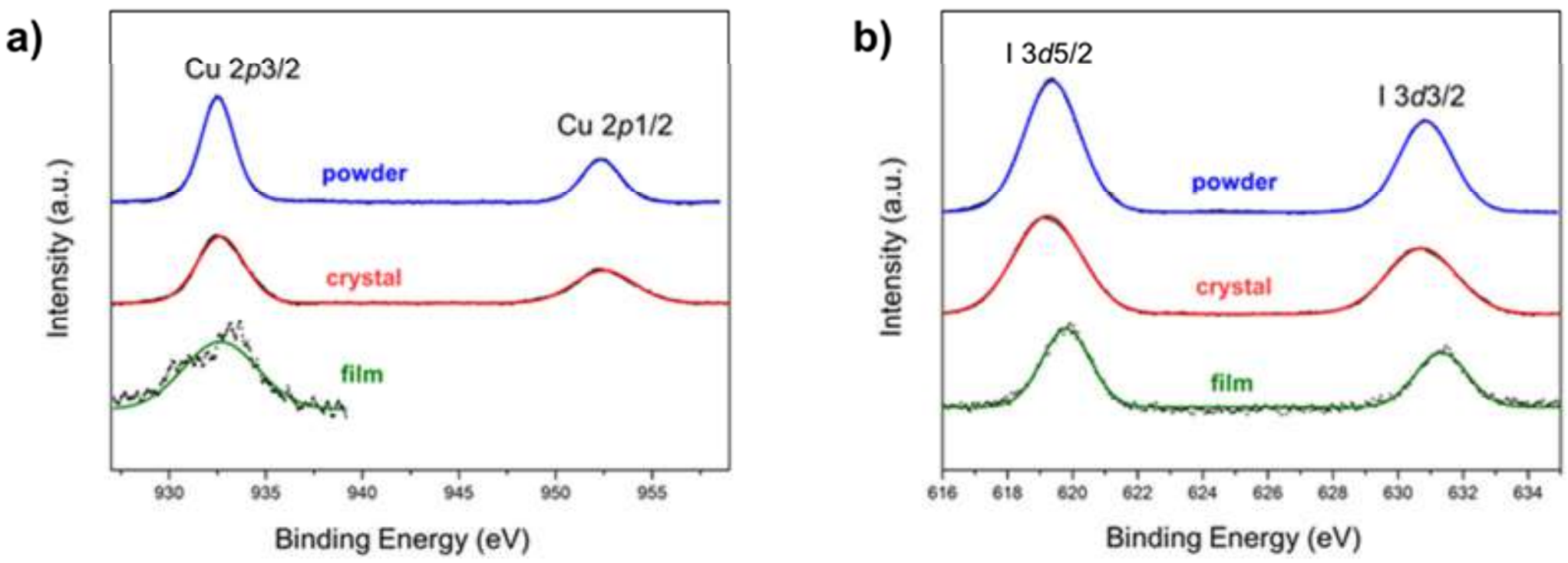

**Figure S3.** XPS copper (a) and iodine (b) spectra of $[Cu_2I_2(TAA)]_n$ powder, crystal and film.

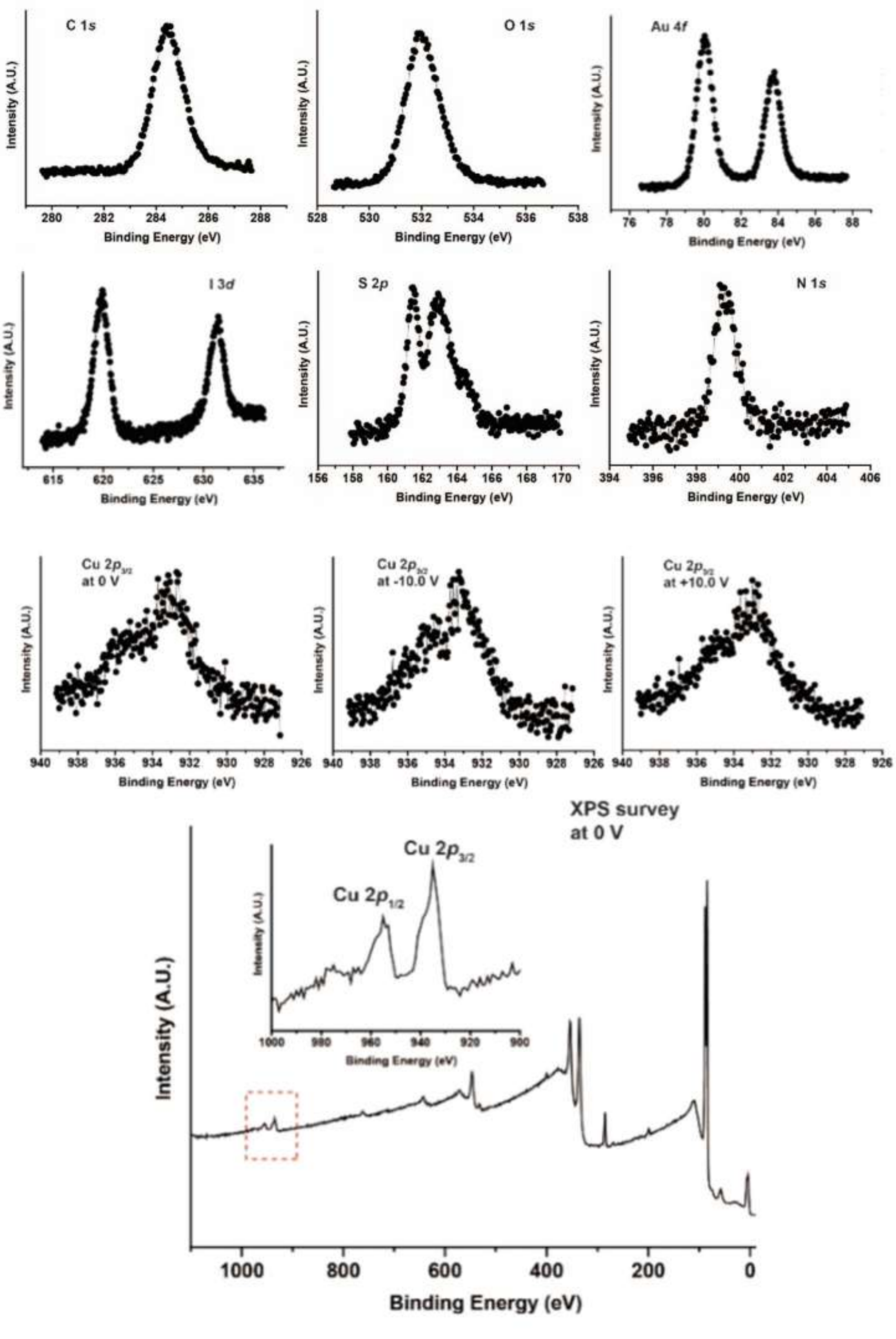

**Figure S4**. High resolution XPS spectra of $[Cu_2I_2(TAA)]_n$ films after applying 10 V characterized by synchrotron XPS. XPS of copper was registered at 0 and ±10 V.

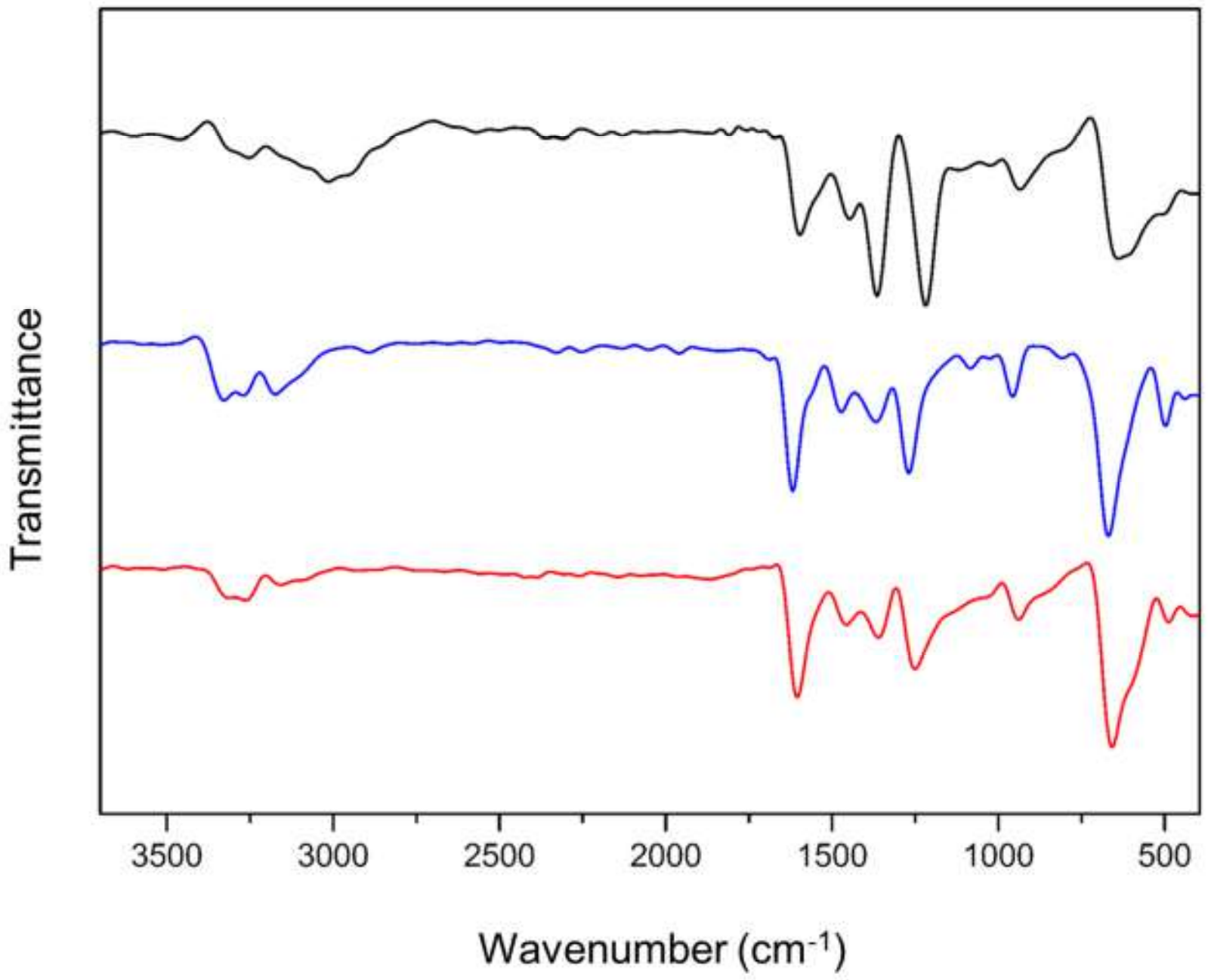

**Figure S5.** FTIR spectra of $[Cu_2I_2(TAA)]_n$ crystals (black), powder (blue) and film (red). The spectra agree with an analogous structure for the $[Cu_2I_2(TAA)]_n$ crystals, powder and film.

For the thermal analysis of $[Cu_2I_2(TAA)]_n$ powder we first obtained the TGA curve and its first derivative. A weight loss of ~15 % was observed from 150 to 350 °C, reaching a plateau for higher temperatures (Figure S6a). The mass loss determined by TGA correlates with the loss of TAA molecules giving rise to a CuI residue (theoretical mass loss = 16 %). This transformation was confirmed by XRPD analysis of the product after heating $[Cu_2I_2(TAA)]_n$ powder at 200 °C for 1 h under inert atmosphere as shown in Figure S6b. Accordingly, we carried out a DSC experiment in this temperature range (from 25 °C to 350 °C) in order to determine the phase transition (T = 170 °C).

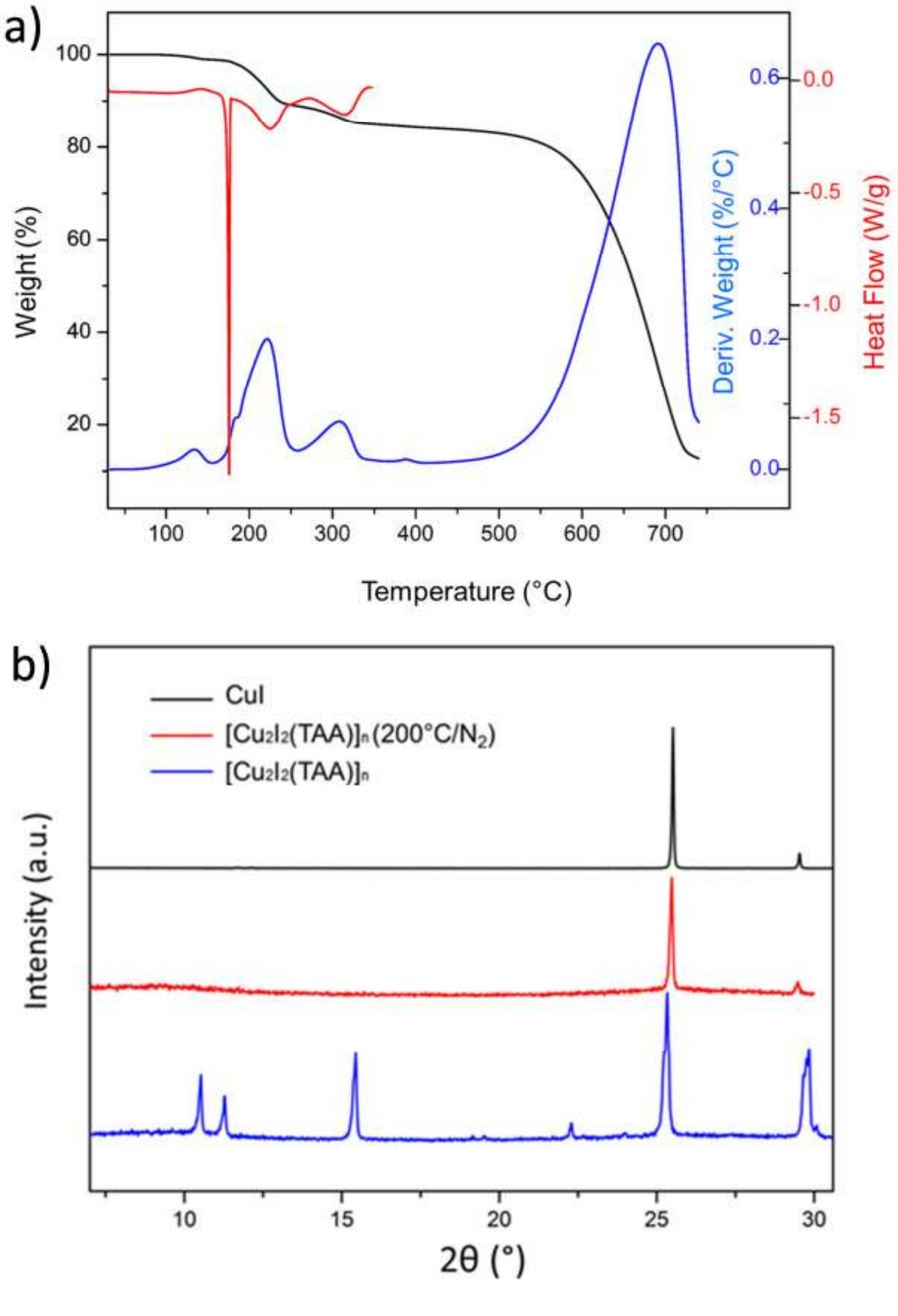

**Figure S6.** a) Thermal gravimetric analysis (TGA) (black), derivative TGA (blue) and differential scanning calorimetry (DSC) (red) of $[Cu_2I_2(TAA)]_n$ powder. b) X-ray powder diffraction of $[Cu_2I_2(TAA)]_n$ powder before (blue) and after (red) heating at 200 °C for 1 h under $N_2$ atmosphere compared to CuI (black).

## SI4. SEM images of films on substrates

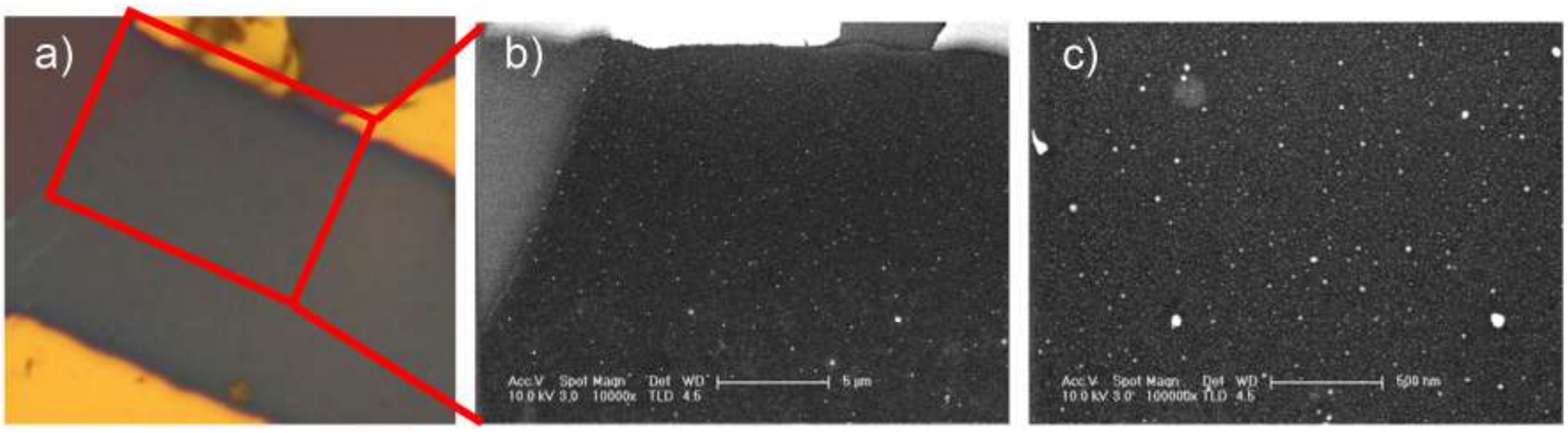

**Figure S7.** a) 30×30 $\mu m^2$ optical image of an 8 nm thick film with gold electrodes on top. b) and c) SEM images with different magnifications of the film in panel a). While SEM images, as well as AFM, show some granular structure they allow discarding a nano-platelet structure as reported previously in similar metal-organic films.

## SI5. Transversal electrical conductivity measurements

The bidimensional behaviour of our films in terms of electrical conductivity was confirmed by measuring transversal conductivity. Films collected on gold surfaces were contacted with $Ga_2O_3$/EGaIn[2] as top electrode (Figure S8a). The measured values of *ca.* $10^{-8}$ S/cm confirmed a tunnel conduction mechanism of the films in the out-of-plane direction (Figure S8b).

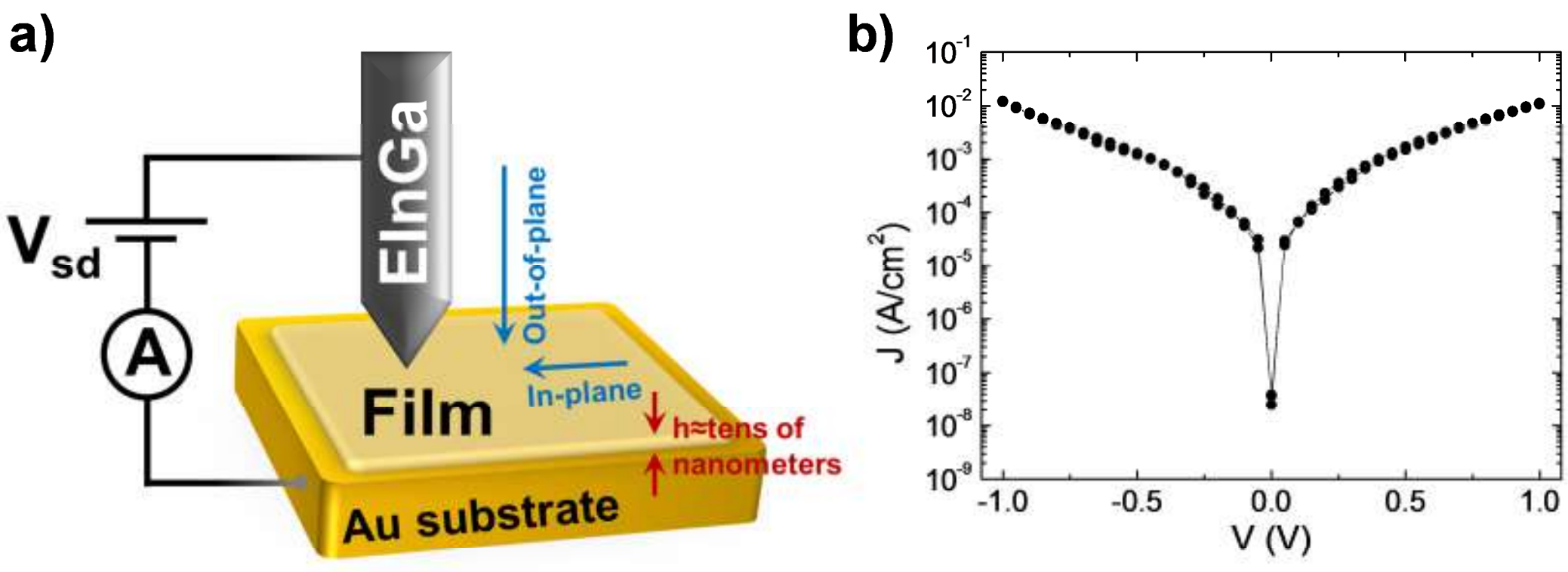

**Figure S8.** a) Experimental scheme used for the measurement of transversal (out-of-plane) conductivity. b) Current density *versus* bias voltage plot from which the transversal conductivity is obtained.

## SI6. KPFM on low conductivity devices

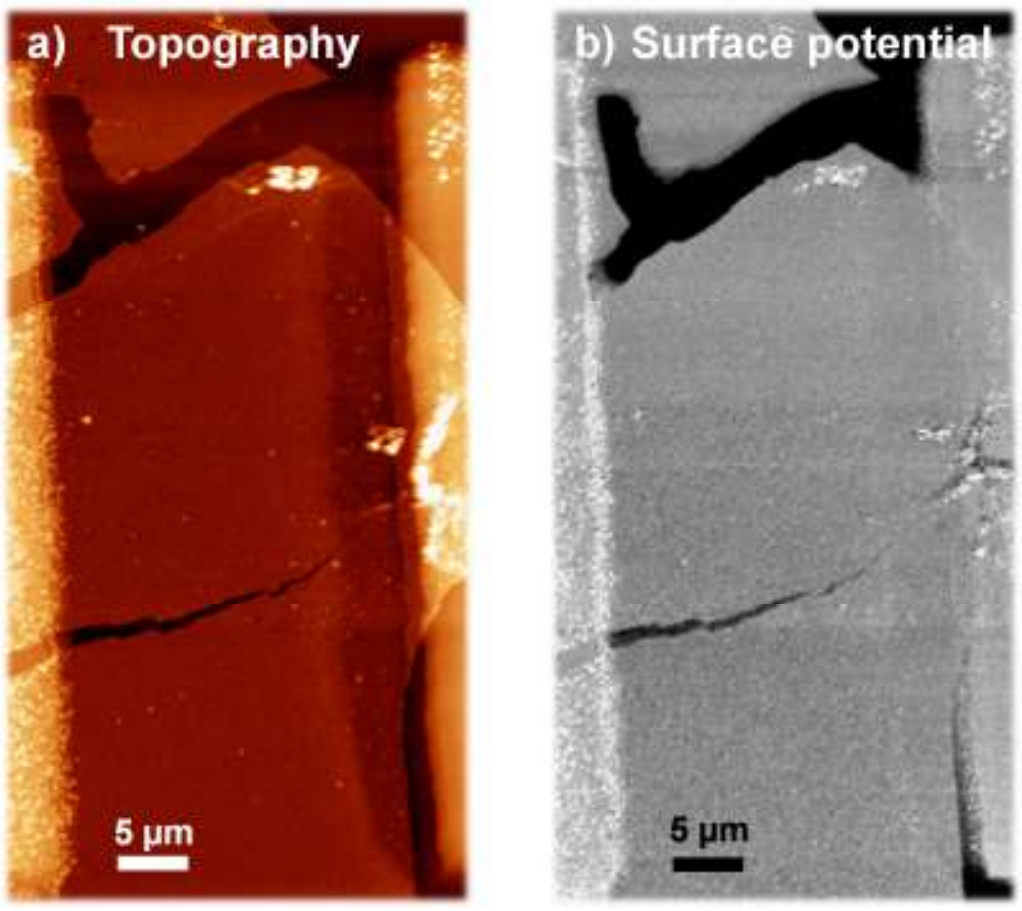

**Figure S9.** Representative a) topography and b) surface potential map of a low conductivity device where the KPFM image appears featureless.

## SI7. Grain boundary migration

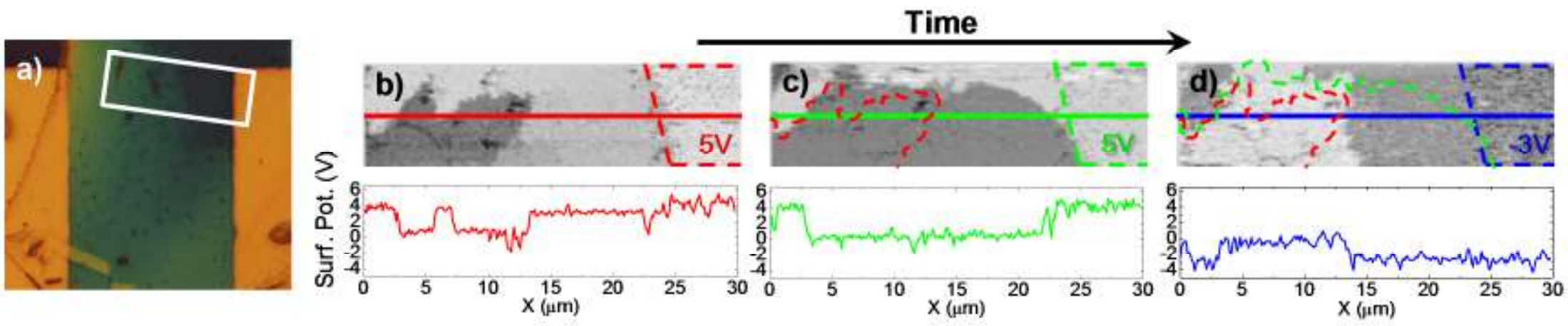

**Figure S10.** a) Optical image of the device where subsequent images were acquired. Image size: 43 × 38 µm$^2$. b, c) Surface potential maps of the region of the device marked in panel a) while 5 V were applied to the right electrode (electrode is indicated by dotted lines). The left electrode is located several microns to the left of each image. It was not scanned in order to minimize the image acquisition time. The time lapse between panels b) and c) is 3 minutes. d) Surface potential map of the same region with -3 V applied to the right electrode. The time between panels c) and d) was 3 minutes. Lower panels are the corresponding profiles along the lines in each image. In c) and d) the borders of different surface potential regions of the previous images are marked with dashed coloured lines. The topography of the studied area remains the same during all the KPFM measurements.

## SI8. *IV* curve fitting to Space Charge Limited Current (SCLC) model

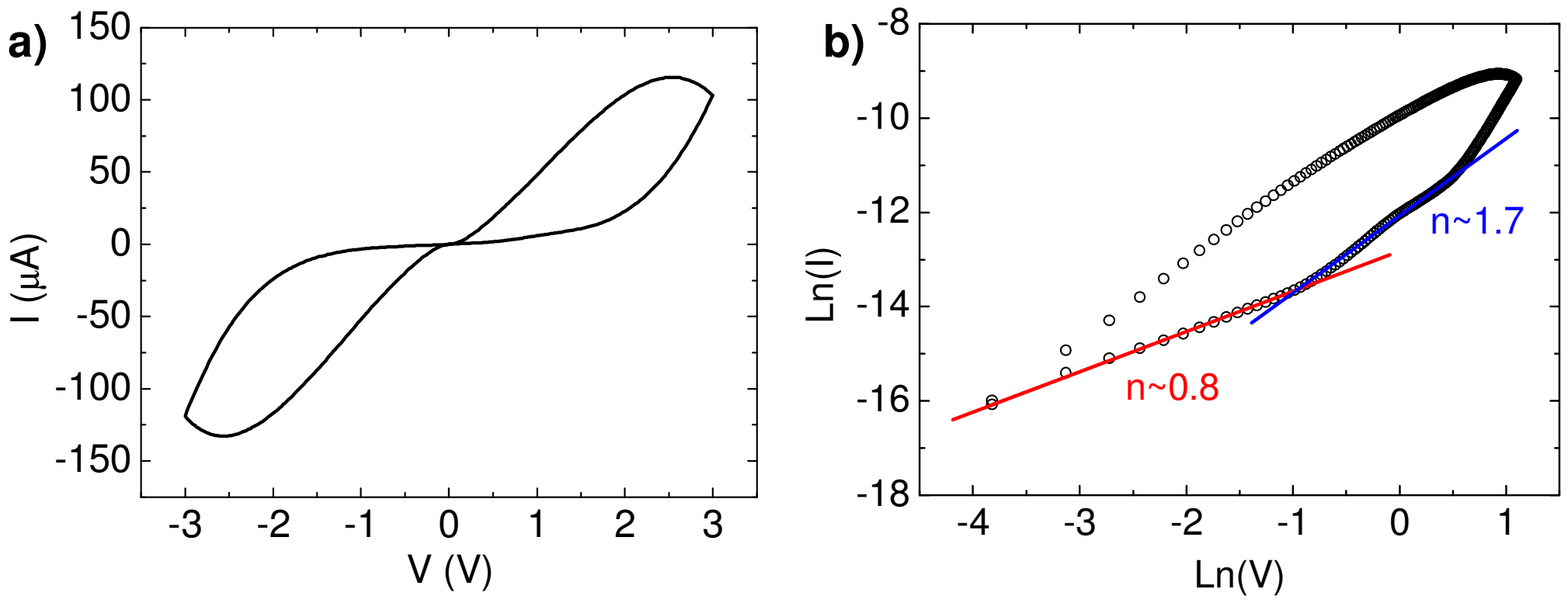

**Figure S11.** a) *IV* curve of a representative film acquired at 0.062 V/s. b) The Ln-Ln representation of the positive-bias loop of the *IV* curve plotted in a) together with linear fittings showing the slopes (exponent in $I \propto V^n$) of two characteristic regions of typical SCLC:[3] an ohmic part at low voltages and a second region with $I \propto V^2$ where the traps begin to be filled with the injected carriers. In the third region with $n > 2$ all traps are filled up, so the subsequently injected carriers can move in the dielectric film causing the subsequent current increasing.